\documentclass[10pt, aps]{revtex4}
\usepackage{amssymb}
\usepackage{latexsym}
\usepackage{epsfig}
\usepackage{float}
\usepackage{graphicx, epsfig, color}
\usepackage{graphicx}
\usepackage{subfigure}
\usepackage{amsmath}
\usepackage{hyperref}

\begin{document}
\title{\textbf{Effect of anomalous magnetic moment on the chiral transition at zero temperature in a strong magnetic field}}
\author{Rui He and Xin-Jian Wen\footnote{wenxj@sxu.edu.cn} }
\affiliation{ Institute of Theoretical Physics, State Key Laboratory of Quantum Optics and Quantum Optics Devices,Shanxi
University, Taiyuan, Shanxi 030006, China}

\begin{abstract}
The effect of the anomalous magnetic moment (AMM) on the chiral
restoration is investigated at zero temperature in the strong
magnetic fields with the vacuum magnetic regularization scheme. It
is shown that the chiral restoration diagram sensitively depends on
the AMM in the ultrastrong magnetic fields. In our work, the
parametrization of AMM is employed as proportional to the square of
the chiral condensate. The critical chemical potential is found to
decrease linearly by the increasing coefficient in the AMM scale. At
a smaller scale of the AMM, the critical chemical potential could go
down and then grow up as the magnetic field increases. But at a
larger scale, the magnetic catalysis on the critical chemical
potential would not happen anymore.
\end{abstract}



\maketitle

\section{Introduction}
Quantum chromodynamics(QCD) is a basic theory to study the strong
interaction between quarks and gluons. It has two main striking
features: asymptotic freedom and color confinement
\cite{Gross:1973ju}. The study of the QCD phase diagram in the
temperature-density plane is a topic that has attracted much
attention over many years. Additionally, the QCD phase diagram also
depends on external parameters, such as the presence of strong
magnetic fields and high densities, which are interesting to
investigate from both experimental and theoretical points of view.
It is well known that, heavy ion collisions can produce a very
strong magnetic field and the order of magnitude is up to about
$eB\sim10^{19}$ Gauss. In the astrophysical environment, strong
magnetic fields still exist in the interior of magnetars
\cite{Duncan:1992hi, Thompson:1993hn}. In the peripheral collisions
of nuclei, extremely larger magnetic field up to $10^{18}$ or higher
value can be generated \cite{
Voronyuk:2011jd,Bzdak:2011yy,Deng:2012pc}. Theoretically, the
maximum strengths of the order $10^{20}$ Gauss in the interior of
stars are proposed by an application of the virial theorem
\cite{Kharzeev:2007jp,Skokov:2009qp}, and even higher fields could
be generated during the electroweak phase transition in the early
Universe \cite{Vachaspati:1991nm, Campanelli:2013mea}.

In recent studies, the chiral phase transition and the equation of
state of dense matter were explored in the strong magnetic field
\cite{Lai:2000at,Miransky:2015ava,Wen:2012jw, Wen:2013yra,
Menezes:2015fla,Ferrer:2021vuy,Wen:2021mgm}. Especially, it is known
that the magnetic catalysis (MC) plays as an important phenomenon,
where a magnetic field enhances the spontaneous chiral symmetry
breakdown. The more general results state that a constant magnetic
field leads to the generation of a fermion dynamical mass
\cite{Miransky:2002rp,Gusynin:1994re, Gusynin:1994xp}. However, in
the region close to the (pseudo) critical temperature, the inverse
magnetic catalysis effect (IMC) is proposed by the lattice QCD
result \cite{ DElia:2011koc}. The finite background magnetic field
leads to the breaking of the chiral symmetry and triggers the
production of quark anomalous magnetic moments (AMM)
\cite{Chang:2010hb,Bicudo:1998qb}. In literature, the AMM is
originally found in the weak-field region, and the Schwinger
linear-in-$B$ ansatz for the AMM of quarks is widely considered
\cite{Fayazbakhsh:2014mca}. In the strong magnetic field region, the
AMM from the one-loop fermion self-energy depends on the Landau
level and decreases with it \cite{Ferrer:2015wca}. The fact of the
dynamic generation of AMM is mainly suggested due to the lowest
Landau level (LLL) effect \cite{Schwinger:1948iu}. Recently, a quark
AMM proportional to the square of chiral condensate
($\kappa_u=\kappa_d=\upsilon\sigma^2$) was suggested to produce
results of chiral condensate as functions of the temperature and the
magnetic field in good agreement with the lattice result
\cite{Kawaguchi:2022dbq}. The AMM was expected to play an important
role to induce the IMC effect around the critical temperature
\cite{Xu:2020yag,Wen:2021mgm}.

The Nambu-Jona-Lasinio (NJL) model was first proposed as a low
energy effective theory for QCD to describe nucleons and mesons. It
was successfully developed to investigate the QCD chiral symmetry
and vacuum spontaneous breakdown at finite density and/or
temperature in a strong magnetic field. However, the four-fermion
interaction in the model leads to the nonrenormalization of the NJL
model, so a proper regularization scheme is needed to avoid
ultraviolet divergences. The familiar regularization schemes are
Pauli-Villars scheme \cite{Mao:2018dqe, Mao:2016fha,
Chaudhuri:2021skc}, the vacuum magnetic regularization scheme(VMR)
\cite{Avancini:2020xqe, Tavares:2021fik, Farias:2021fci}, the
magnetic field independent regularization scheme (MFIR)
\cite{Aguirre:2020tiy}, and non-MFIR scheme. Unfortunately, the
non-MFIR schemes will produce nonphysical oscillation behavior in
chiral quark condensate or tachyonic neutral pion masses
\cite{Chaudhuri:2019lbw}. The MFIR scheme and the VMR scheme are
helpful to extract physical content from the vacuum of the strong
interaction affected by a magnetic field. In this work, we will
employ the VMR scheme to deal with the divergencies in the
thermodynamic potential and discuss the influence of AMM on the
chiral phase transition at finite densities. The effect of AMM on
the chiral restoration at finite temperature has been studied by the
VMR scheme with convincing results \cite{Farias:2021fci}. Our aim
focuses on the possible AMM scale dependent on the chiral quark
condensate and its effect on the critical chemical potential in a
strong magnetic field.

The paper is organized as follows. In Sec. II, we present the
thermodynamics of the two-flavor NJL model with nonzero AMM in a
strong magnetic field. In Sec. III, the numerical results are shown
with a detailed investigation on the influence of AMM on chiral
phase transition. The last section is a short summary.

\section{Thermodynamics of the SU(2) NJL model at zero temperature}
\label{sec2}

In the SU(2) version of the NJL model under a strong magnetic field,
the Lagrangian density of the two-flavor NJL model is given by
\begin{equation}
{\mathcal{L}}_{NJL}=\bar{\psi}(i/\kern-0.7em D -m +\frac{1}{2} \hat{a} \sigma^{\mu\nu}F_{\mu\nu}   )\psi +G[(\bar{\psi}%
\psi )^{2}+(\bar{\psi}i\gamma _{5}\vec{\tau}\psi )^{2}].
\end{equation}
where $\psi $ represents a flavor isodoublet ($u$ and $d$ quarks) and $%
\vec{\tau}$ is the isospin Pauli matrix. The coupling of the quarks
to the electromagnetic field is introduced by the covariant
derivative $/\kern-0.7em D \sim \gamma^\mu$$D_\mu$ and
$D_\mu=\partial_\mu-i e \hat{Q} A_\mu$. The charge matrix is given
by $\hat{Q}\equiv \mathrm{diag}(q_u, q_d)=\mathrm{diag}(2/3, -1/3)$.
The Abelian gauge field $A_\mu$ stands for the external magnetic
field $B$ aligned along the z-direction. The AMM is introduced by
the $\sigma^{\mu\nu}=i[\gamma^\mu, \gamma^\nu]/2$ coupling with
electromagnetic field strength $F^{\mu\nu}=\partial^\mu
A^\nu-\partial^\nu A^\mu$. The matrix tensor used in this work is
$g^{\mu\nu}$= $\mathrm{diag}(1, -1, -1, -1)$. The factor $\hat{a}
=\hat{Q}\hat{\kappa}$, where $\hat{\kappa}$ =
$\mathrm{diag}(\kappa_u, \kappa_d)$, is a 2 $\times$ 2 matrix in the
flavor space; here $\kappa_i$ are AMM of the quarks, The more recent
results suggested that the proper form of AMM would change with the
chiral condensate, since it involves the behavior related to the
condensate \cite{Kawaguchi:2022dbq}.

By expanding $ \bar{\psi}\psi$ around the quark condensate $\langle
\bar{\psi}\psi \rangle$ and dropping the quadratic term of the
fluctuation, one can get the mean-field approximation
$\left(\bar{\psi}\psi\right)^2\approx2\langle \bar{\psi}\psi
\rangle\left(\bar{\psi}\psi\right)-\langle \bar{\psi}\psi
\rangle^2$. The dynamical quark mass is given by
\begin{equation}
M_i=m-2G\langle \bar{\psi}\psi \rangle , \label{eq:gap}
\end{equation}
where the quark condensates include $u$ and $d$ quark contributions
as $\langle \bar{\psi}\psi \rangle \equiv \sigma=\sum_{i=u, d}\sigma
_{i}$. The dynamical mass depends on both flavors condensates.
Therefore, the same mass $ M_{u}=M_{d}=M$ is available for $u$ and
$d$ quarks. The contribution from the $i$ flavor quark is
\cite{Farias:2021fci}
\begin{equation}
\sigma_{i}=\sigma _{i}^{\mathrm{vac}}+\sigma_i ^{%
    \mathrm{field}}+\sigma _{i}^{%
\mathrm{mag}} +\sigma_i^\mathrm{med}. \label{eq:condensate}
\end{equation}
The terms $\sigma_i ^{\mathrm{vac}}$, $\sigma_i ^{%
\mathrm{field}}$ and $\sigma_i ^{%
\mathrm{mag}}$ represent the vacuum, the field, and the magnetic
field to the quark condensation, respectively. The regularized
vacuum contribution reads
\begin{eqnarray}
\sigma_i ^{\mathrm{vac}}= -\frac{M N_c}{2\pi ^{2}}\{\Lambda\epsilon
_{i}(\Lambda)-K_{0i}^{2}\ln[\frac{\Lambda+\epsilon
_{i}(\Lambda)}{K_{0i}}]\},\label{eq:gap1}
\end{eqnarray}
where a 3D sharp cutoff $\Lambda$ of the momentum is employed. The
definitions $K_{0i}=\sqrt{M^2+\kappa_i^2 B_i^2}$ and $\epsilon
_{i}^{2}(\Lambda)= K_{0i}^{2}+\Lambda ^{2}$ are adopted to include
the AMM with the parameter $B_i$ defined as $B_i=q_ieB$
\cite{Farias:2021fci}. The finite magnetic field-dependent
contributions are given by
\begin{eqnarray}
\sigma_i ^{\mathrm{field}} &=&-\frac{M N_c B_i^2}{24\pi
^{2}}\frac{[3(\alpha_i+1)^2-1]}{K_{0i}^{2}}, \label{eq:gap2} \\
\sigma_i^{\mathrm{mag}} &=& -\frac{M N_c}{4\pi
^{2}}\int_{0}^{\infty}\frac{ds}{s^2}e^{-sK_{0i}^{2}}\times\{\frac{B_is\cosh[(\alpha_i+1)B_is]}{\sinh(B_is)}-1-\frac{1}{6}\left[3(\alpha_i+1)^2-1\right](B_is)^2\}
\label{eq:gap3}
\end{eqnarray}
with the notation $\alpha_i=2M\kappa_i$. $\sigma_i^\mathrm{med}$ is
contribution of medium at zero temperature, given by the following
expression
\begin{eqnarray}
\begin{aligned}
\sigma_i^\mathrm{med}=
\frac{MN_c\left|B_i\right|}{2\pi^2}\sum_{n=0}^{n_\mathrm{max}}\sum_{s=\pm1}\left(1-\frac{sT_i}{M_{nis}}\right)\ln
\left(\frac{\mu_i+p_F}{M_{nis}-sT_i}\right), \label{eq:gap4}
\end{aligned}
\end{eqnarray}
where we have adopted the Landau-level induced energy eigenvalue
\begin{eqnarray}
M_{nis}&=&\sqrt{(2n+1-s\frac{q_i}{|q_i|})|B_i|+M^2},
\end{eqnarray}
and the longitudinal Fermi momentum
\begin{equation}
p_F=\sqrt{\mu_i^2-(\sqrt{(2n+1-s\frac{q_i}{|q_i|})|B_i|+M^2}-sT_i)^2},
\end{equation}
Due to the requirement $p_F\ge0$, one can get the maximum
Landau-level number
\begin{equation}
n_\mathrm{max} =
\mathrm{Floor}\left[\frac{1}{2}\left(\frac{(sT_i+\mu_i)^{2}-M^{2}}{\left|B_i\right|}+s\frac{q_i}{\left|q_i\right|}-1\right)\right],
\end{equation}
where $T_i=\kappa_iB_i$ includes the AMM according to the Schwinger
linear ansatz
 \cite{Schwinger:1948iu}, and
$s=\pm 1$ stands for the spin of the quark. The AMM separates the
energies of the up and down spins in the LLs ($n\neq 0$), in
addition to the LLL ($n=0$).

In the VMR scheme, the NJL thermodynamic potential density can be
written as \cite{Farias:2021fci}:
\begin{eqnarray}
\Omega_i=\frac{(M-m)^2}{4G}+\sum_{i=u, d}\Omega_i ^{\mathrm{vac}}+\Omega_i ^{\mathrm{field}}+\Omega_i ^{\mathrm{mag}}+\Omega_i^\mathrm{med},
\end{eqnarray}
The contributions $\Omega_i ^{\mathrm{vac}}$ and $\Omega_i
^{\mathrm{field}}$ must be regularized and the following expressions
are given by
\begin{eqnarray}
\Omega_i ^{\mathrm{vac}}&=&-\frac{N_{c}}{8\pi
^{2}}\begin{Bmatrix}\Lambda [\Lambda ^{2}+\epsilon
_{i}^{2}(\Lambda)]\epsilon _{i}(\Lambda)-
K_{0i}^{4}\ln[\frac{\Lambda
+\epsilon_{i}(\Lambda)}{K_{0i}}]\end{Bmatrix}, \\
\Omega_i^{\mathrm{field}}&=&-\frac{N_{c}B_i^2}{48\pi
^{2}}\left[3(\alpha_i+1)^2-1\right]\ln\frac{K_{0i}^{2}}{\Lambda
^{2}},
\end{eqnarray}
Because of the ultraviolet divergence, we still use the 3D sharp
cutoff scheme to regularize the vacuum term. The
$\Omega_i^\mathrm{mag}$ is the magnetic field contributions
\cite{Farias:2021fci},
\begin{eqnarray}
\begin{aligned}
\Omega_i^\mathrm{mag}=& \frac{N_{c}}{8\pi
^{2}}\int_{0}^{\infty}\frac{ds}{s^3}e^{-sK_{0i}^{2}}\begin{Bmatrix}\frac{B_is\cosh[(\alpha_i+1)B_is]}{\sinh(B_is)}-1-\frac{1}{6}\left[3(\alpha_i+1)^2-1\right](B_is)^2\end{Bmatrix}.
\end{aligned}
\end{eqnarray}
The contribution of the medium at zero temperature
$\Omega_i^\mathrm{med}$ is,
\begin{equation}
\Omega_i^\mathrm{med}=-\frac{N_c\left|B_i\right|}{4\pi^2}\sum_{n=0}^{n_\mathrm{max}}\sum_{s=\pm1}\left[
p_F \mu_i - (M_{nis}-sT_i)^2\times\ln
\left(\frac{\mu_i+p_F}{M_{nis}-sT_i}\right)\right].
\end{equation}
From the thermodynamic potential, the quark number density is easily
evaluated as follows:
\begin{eqnarray}
\rho_i=\frac{N_c|B_i|}{2\pi^2} \sum_{n=0}^{n_\mathrm{max}}
\sum_{s=\pm}\sqrt{\mu_i^2-(M_{nis}-sT_i)^2}.
\end{eqnarray}

For completeness, at the zero AMM, the replacements of
$K_{0,i}\rightarrow M$, $\alpha_i\rightarrow 0$, and $T_i\rightarrow
0$ should be done in Eqs.~(\ref{eq:gap1})$-$(\ref{eq:gap4}) to get
the gap equation. It has been verified that the thermodynamic
potential $\Omega_\mathrm{VMR}$ can be taken as a modification of
the potential $\Omega_\mathrm{MFIR}$ by an additional term, which is
independent on the effective mass $M$ \cite{Avancini:2020xqe}. As a
consequence, the gap equation in VMR scheme is compatible with the
result from MFIR scheme in Ref. \cite{Menezes:2008qt}. Specially,
the vacuum effective mass $M$ without AMM is still given by
\begin{eqnarray}\frac{M-m}{2G}&=&-\frac{M N_c N_f}{2\pi ^{2}}\Big\{\Lambda\epsilon
_{i}(\Lambda)-M^{2}\ln[\frac{\Lambda+\epsilon
_{i}(\Lambda)}{M}]\Big\}
\nonumber \\
&& +\sum_{i=u,d}\frac{MN_c |B_i|}{2\pi^2}\Big\{
\ln[\Gamma(x_i)]-\frac{1}{2}\ln(2\pi)+x_i-\frac{1}{2}(2
x_i-1)\ln(x_i) \Big\},
\end{eqnarray}
where $x_i= M^2/|2B_i|$ is defined \cite{Menezes:2008qt}. The
detailed derivation of the equivalence of the gap equations in two
schemes can be found in Ref.~\cite{Avancini:2020xqe}.

\section{Numerical result and discussion}

In this section, the AMM effect at high densities is studied in
terms of the chiral phase transition in the strong magnetic field.
In the present calculation, the following parameters are adopted;
$m_u=m_d=4.548$ MeV, $\Lambda=719.23$MeV, and $G=1.954/\Lambda^2$.
In Ref. \cite{Kawaguchi:2022dbq}, the three forms of the AMM
($\kappa=$constant, $\kappa=\upsilon \sigma$, and $\kappa=\upsilon
\sigma^2$) were compared and discussed. The scale proportional to
the square of the condensate was considered as the practicable
effective form to describe the thermomagnetic properties of QCD. The
AMMs for $u$ and $d$ quarks in our work are adopted as
$\kappa_u=\kappa_d=\upsilon\sigma^2$. The opposite signs
$\kappa_u=-\kappa_d=\upsilon\sigma^2$ for possible negative
contribution of $d$ quarks are considered for comparison The quark
dynamical mass as a function of the chemical potential in different
magnetic fields can be obtained by solving the gap equation
Eq.~(\ref{eq:gap}), Then we can analyze the effect of AMM on the
critical chemical potential for the first-order phase transition. In
the calculations, we assume the isospin symmetric case meaning that
the chemical potentials are equal $\mu_u=\mu_d=\mu$ for $u$ and $d$
quarks.

\begin{figure}[H]
    \centering
    \includegraphics[width=8cm]{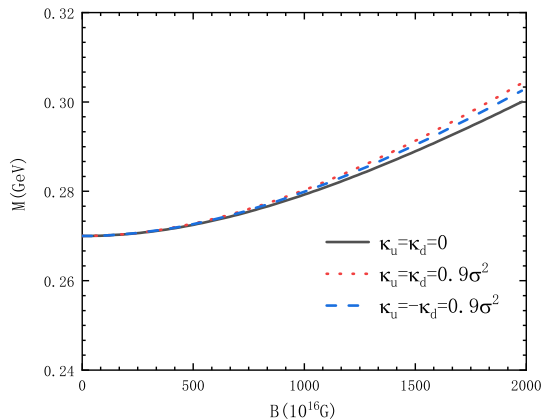}
    \caption{\label{fig:wide}The behavior of the effective quark mass in vacuum state with and without AMMs as a function of the magnetic
    field.}\label{fig:vac}
\end{figure}
In Fig.~\ref{fig:vac}, the vacuum mass is shown as a function of the
magnetic field intensity. The three AMMs $\kappa=0$,
$\kappa_u=\kappa_d=0.9 \sigma^2$, and
$\kappa_u=-\kappa_d=0.9~\sigma^2$ are marked by the black solid, the
red dotted, and the blue dashed lines respectively. The behavior of
the effective quark mass enhanced by the magnetic field is
consistent with the result pointed in Ref. \cite{Menezes:2008qt}.
From the figure, it is clearly seen that the growing behavior of the
effective mass is relatively slightly enhanced by the nonzero AMM,
where the larger magnitude of the magnetic field facilitates the
binding of the quark and antiquark.

\begin{figure}[h]
\centering
 \includegraphics[width=8cm]{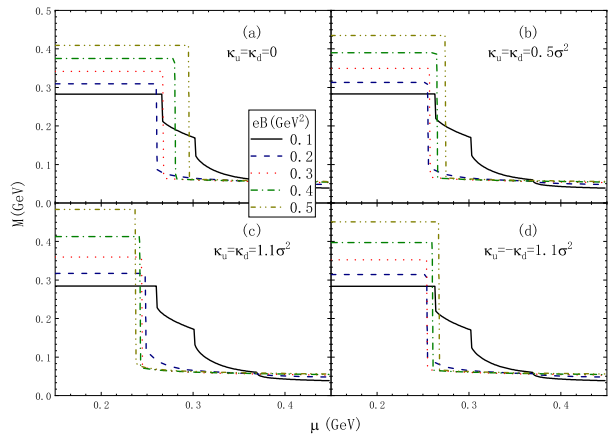}
\caption{\label{fig:wide}The quark effective mass as a function of
chemical potentials for different magnetic fields at the AMMs (a)
$\kappa_u=\kappa_d=0$, (b) $\kappa_u=\kappa_d=0.5 \sigma^2$, (c)
$\kappa_u=\kappa_d=1.1 \sigma^2$, and (d) $\kappa_u=-\kappa_d=1.1
\sigma^2$.}\label{fig:mass}
\end{figure}

In Fig. \ref{fig:mass}, we show the dynamical quark effective mass
as functions of the chemical potential at the different AMMs in four
panels. The magnetic fields are adopted from 0.1 GeV$^2$ to 0.5
GeV$^2$, which are clearly marked by the curves from bottom to top
in the vacuum state. The descending behavior of quark mass occurs in
the chiral restoration process, which is not a smooth slope but a
sudden drop denoting the first-order transition. In the magnetic
field of 0.2 GeV$^2$ or even higher order, the quarks only occupy
the LLL to give rise to a single first-order phase transition. For
the weaker magnetic field $eB=0.1$ GeV$^2$ marked by the solid
lines, the maximum Landau levels for $u$ and $d$ quarks are
mismatched \cite{Wang:2022xxp}. In Table~\ref{tab:level}, the
quantum number of maximum Landau levels $n_{u,max}$ and $n_{d,max}$
is listed in the chemical potential $\mu$ range from vacuum to the
chiral restoration. At $\mu>267$~MeV for $\kappa=0$ or $\mu>263$~MeV
for $\kappa=0.5 \sigma^2$, the LLL (n=0) occurs for $u$- and $d$-
quarks. When the chemical potential increases up to about 302 MeV,
the energy level is excited to a higher level for $d$ quarks, while
the $u$ quarks still lie in the LLL. The quarks occupation would
influence the condensate of quark and antiquarks. As a consequence,
there are two first-order transitions in the region of densities.
Comparing the panels (a) (b) and (d) in Fig.~\ref{fig:mass}, it can
be found that the AMM would result in the critical chemical
potential moving to larger value with the increasing magnetic field
in the region of $eB>0.2$GeV$^2$. It is characterized by the
so-called MC effect and is compatible to the conclusion in the
absence of AMM \cite{Garcia:2013eaa, Wang:2022xxp}. On the contrary
in panel (c), the larger coefficient with the same sign of
$\kappa_u$ and $\kappa_d$ would always produce the IMC effect, which
is characterized by the decrease of the critical chemical potential
as the magnetic field increases. By comparing the vertical axis on
four panels, it can be found that the AMM has poor impact on the
vacuum mass of quarks in the weak magnetic field. When the magnetic
field becomes strong enough, the enhancement of the vacuum chiral
condensate by the AMM becomes more evident. The effect of AMM is
agreement with those obtained at finite temperature in Ref.
\cite{Farias:2021fci}. It is concluded that the scale and sign of
AMM would have significant effect on the realization of the MC and
IMC.

\begin{table}[htbp]
\centering \caption{\label{tab:level} The quantum number of Landau
Levels occupied by quarks for the magnetic field $eB=0.1$ GeV$^2$ at
different AMMs. The number $``0"$ means the LLL.}
\begin{tabular}{cccccc}
\hline \hline AMM ($\kappa/\sigma^2$)& \ $\mu$ (MeV)  \ &\ $n_{u,\mathrm{max}}$ \  &\  $n_{d,\mathrm{max}}$ \  \\
\hline
  &  \ \ 0$\sim$ 267 & No  & No   \\
0 &  267$\sim$ 303 & 0  & 0   \\
 &  303$\sim$ 369  & 0  & 1   \\
 \hline
  &  \ \ 0$\sim$ 263 & No  & No    \\
0.5 &   263$\sim$ 302 & 0  & 0   \\
 &   302$\sim$ 369 & 0  & 1   \\
  \hline \hline
\end{tabular}
\end{table}

\begin{figure}[htbp]
    \centering
    \includegraphics[width=8cm]{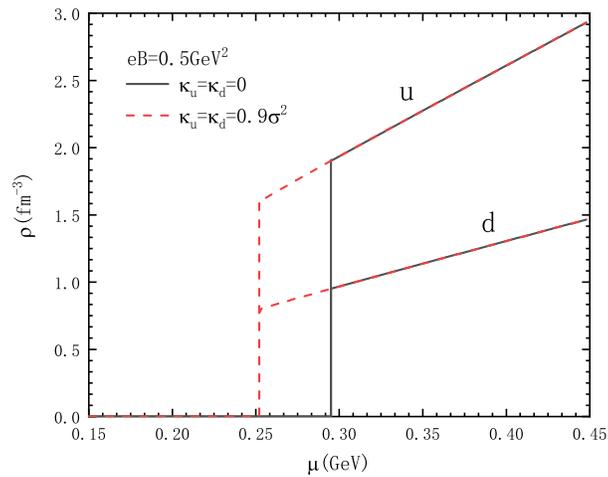}
    \caption{The quark number density with and without AMMs as a function of the chemical potential.}\label{fig:rho}
\end{figure}

In the strong magnetic field, the charged quark would be ruled in
the Landau levels. The $u$ and $d$ quarks will have different
densities with the same chemical potential due to their different
electric chargers. In Fig. \ref{fig:rho}, the quark density $\rho_u$
and $\rho_d$ are shown at the magnetic field $eB=0.5$~GeV$^2$. The
presence of the AMM marked by the dashed lines affects not the
number density but the position of the occurrence of quarks. The
degenerate factor proportional to $|q_iB|$ is responsible for the
relation $\rho_u>\rho_d$, where the absolute value of the electric
charge of $u$ quarks is larger than that of $d$ quarks.

\begin{figure}[htbp]
    \centering
    \includegraphics[width=8cm]{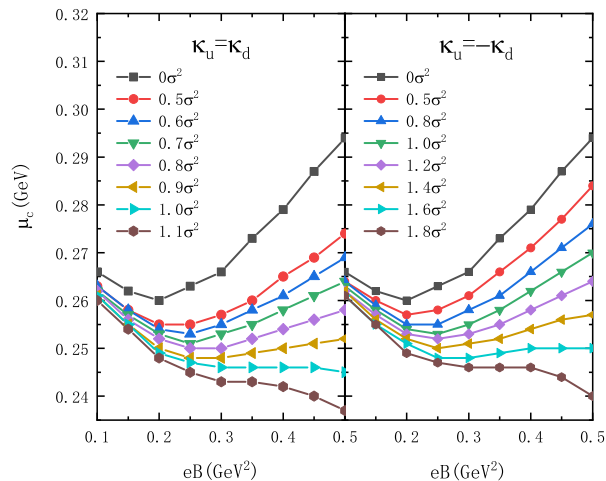}
    \caption{The behavior of the critical chemical potential for different AMM scales as a function of the magnetic
    field.}\label{fig:mu-eB}
\end{figure}
In Fig. \ref{fig:mu-eB}, the critical chemical potential for the
chiral restoration is shown as a function of the magnetic field at
the AMMs: $\kappa_u=\kappa_d$ on left panel and $\kappa_u=-\kappa_d$
on right panel. The different coefficients of the AMM proportional
to the square of the chiral condensate are marked on the lines. For
the $\kappa_u=\kappa_d<1.0 \sigma^2$ on left panel, one would see
the behavior of the critical chemical potential $\mu_c$ going down
and up with the increase of the magnetic field. On these
nonmonotonously curves, there exists a special value for the
magnetic field $eB_c$, above which the MC effect is revealed by the
increase of $\mu_c$ with the increase of $eB$. While for the
magnetic fields below the value, the IMC effect would operate
slightly in a short range. But as the coefficient of AMM increases
up to 1.0, the nonmonotonous behavior disappears and the IMC effect
would be always realized in the whole range of the magnetic field.
If the sign of the AMM for $u$ and $d$ quarks is opposite, namely,
$\kappa_u=-\kappa_d=\upsilon \sigma^2$, a similar behavior of
$\mu_c$ is obtained in addition to that the much larger coefficient
is available at the same $\mu_c$ on right panel. Generally speaking,
for two cases at any fixed magnetic field, it can be concluded that
the larger coefficient of AMM would lead to the smaller $\mu_c$ for
the chiral transition.

\begin{figure}[htbp]
    \centering
    \includegraphics[width=8.0cm]{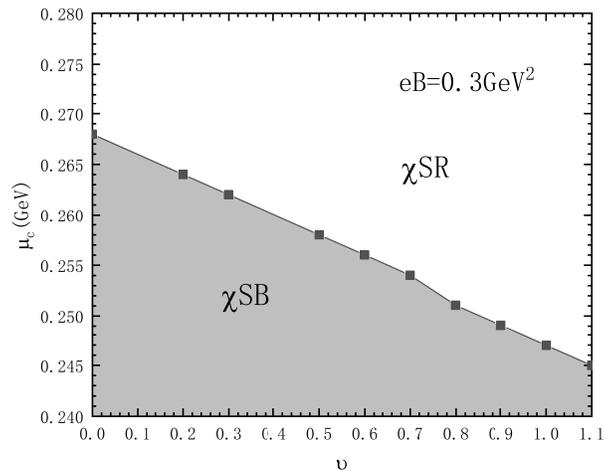}
\caption{\label{fig:wide}The critical chemical potential as a
function of the coefficient $\upsilon$ and the region of $\chi SR$
and $\chi SB$ is separated for the magnetic field
$eB=0.3~\mathrm{GeV}^2$. }\label{fig:u}
\end{figure}

In Fig. \ref{fig:u}, the critical chemical potential $\mu_c$ is
shown as a function of the coefficient of AMM at the magnetic field
$eB=0.3$ GeV$^2$. It is interestingly found that the $\mu_c$ is
nearly a linearly decreasing function of the coefficient of AMM. The
gray part below the critical line denotes the chiral symmetry
breaking ($\mathrm{\chi SB}$). At the chemical potential larger than
$\mu_c$, the chiral symmetry restoration ($\mathrm{\chi SR}$) is
expected to take place in the upper blank region. The AMM with the
coefficient larger than 1.0 will result in a decrease of $\mu_c$
close to $10 \%$ of the original value without AMM.

\begin{figure}[htbp]
    \centering
    \includegraphics[width=8cm]{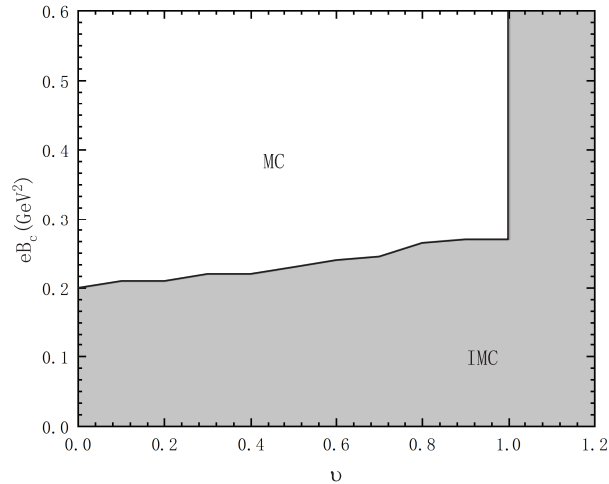}
\caption{\label{fig:wide}The regions of MC and IMC are shown in the
$eB-\upsilon$ plane. }\label{fig:MCIMC}
\end{figure}

The scale of AMM could have a significant impact on the role of the
magnetic field in the chiral restoration. As was discussed above,
there is a critical magnetic field separating the region of MC and
IMC discovered in Fig. \ref{fig:mu-eB}. Now one can continuously
change the coefficient of AMM proportional to the square of chiral
condensate and get the whole diagram of MC and IMC in the
$eB-\upsilon$ plane in Fig. \ref{fig:MCIMC}. The boundary between
the MC and IMC is described by a solid line. For a given scale of
the AMM, the IMC effect can always be realized in a weaker magnetic
field marked by the gray area and the MC effect in a stronger
magnetic field marked by the empty region. But for the coefficient
of the AMM increasing up to 1.0, the larger value of the AMM
$\kappa=1.0 \sigma^2$ is obtained and the IMC region would expand to
the whole area. On the other hand, for a given magnetic field, the
increase of the coefficient of the AMM could make the possibility
for the MC turning to the IMC effect. It can be concluded that the
scale of AMM is crucial to account for the happening of IMC and MC
effects in the chiral restoration.

\section{Summary}
In this paper, we have explored the effect of the AMM on the chiral
restoration at zero temperature to the strong magnetic fields. The
VMR scheme has been used to avoid UV divergence. The chiral
restoration happens with a sudden drop to indicate a first-order
transition at larger densities. We found that the AMM proportional
to the square of the chiral condensate, i.e., $\kappa=\upsilon
\sigma^2$, has a crucial impact on the chiral restoration. Even
though the effect of the AMM can be negligible in the region of the
weak magnetic field, the enhancement of the dynamical vacuum mass is
very sensitive to the AMM as the magnetic field becomes stronger.
The critical chemical potential would slightly decrease with the
increasing coefficient in the scale of the AMM $\kappa=\upsilon
\sigma^2$. The AMM of $\kappa>1.0 \sigma^2$ would give rise to a
decrease of the critical chemical potential up to 10\% of the
original value without AMM.

In the case of small scale of AMM, the critical chemical potential
is a nonmonotonous function of the magnetic field. In the convex
behavior, it is seen that the IMC effect occurs in weaker magnetic
fields and the MC effect occurs in stronger magnetic fields. The
increase of the coefficient $\upsilon$ would turn the MC effect into
IMC effect. For the larger scale near to $\kappa=1.0 \sigma^2$, the
inverse magnetic catalysis region would expand to the whole area in
the $eB-\upsilon$ plane. So it is concluded that the occurrence of
the MC effect, namely, the decrease of the critical chemical
potential with the magnetic field, would constrain an upper limit on
the scale of the AMM. It is expected that our result is instructive
for the investigation on AMM in future experiments.

\acknowledgments{ The authors would like to thank support from the
National Natural Science Foundation of China under the Grants No.
11875181, No. 11705163, and No. 12147215. This work was also
sponsored by the Fund for Shanxi "1331 Project" Key Subjects
Construction. }

\end{document}